\documentclass[proof]{pasj00}
\Received{2013 November 28}
\Accepted{2014 February 7}
\SetRunningHead{Astronomical Society of Japan}{Kamitsukasa et al.}
\usepackage{times}
\usepackage{natbib}
\usepackage{comment}
\usepackage{lscape}
\usepackage{ulem}
\begin{document}
\title{Suzaku Discovery of Fe K-Shell Line from the O-Rich SNR G292.0$+$1.8}
\author{Fumiyoshi \textsc{Kamitsukasa},\altaffilmark{1}
        Katsuji \textsc{Koyama},\altaffilmark{1,2}
        Hiroshi \textsc{Tsunemi},\altaffilmark{1}
        Kiyoshi \textsc{Hayashida},\altaffilmark{1}
        Hiroshi \textsc{Nakajima},\altaffilmark{1}
        Hiroaki \textsc{Takahashi},\altaffilmark{1}
        Shutaro \textsc{Ueda},\altaffilmark{1}
        Koji \textsc{Mori},\altaffilmark{3}
        Satoru \textsc{Katsuda},\altaffilmark{4}
        and Hiroyuki \textsc{Uchida} \altaffilmark{2}}
\altaffiltext{1}{%
   Department of Earth and Space Science, Osaka University, 1-1 Machikaneyama-cho,
   Toyonaka, Osaka 560-0043, Japan}
\email{kamitsukasa@ess.sci.osaka-u.ac.jp}
\altaffiltext{2}{%
   Department of Physics, Graduate School of Science, Kyoto University, 
   Kitashirakawa Oiwake-cho, Sakyo-ku, Kyoto 606-8502, Japan}
\altaffiltext{3}{%
   Department of Applied Physics and Electronic Engineering, Faculty of Engineering, 
   University of Miyazaki, \\1-1 Gakuen Kibanadai-Nishi, Miyazaki 889-2192, Japan}
\altaffiltext{4}{%
   RIKEN (The Institute of Physical and Chemical Research) Nishina Center, 
   2-1 Hirosawa, Wako, Saitama 351-0198, Japan}

 %Do NOT move this preamble from here!

\maketitle
\begin{abstract}

We report the Suzaku/XIS results of the Galactic oxygen-rich supernova remnant (SNR), G292.0$+$1.8, a remnant of a core-collapse supernova. The X-ray spectrum of G292.0$+$1.8 consists of two type plasmas, one is in collisional ionization equilibrium (CIE) and the other is in non-equilibrium ionization (NEI). The CIE plasma has nearly solar abundances, and hence would be originated from the circumstellar and interstellar mediums. The NEI plasma has super-solar abundances, and the abundance pattern indicates that the plasma originates from the supernova ejecta with a main sequence of 30--35\,{\it{M}}$_{\odot}$.  Iron K-shell line at energy of 6.6\,keV is detected for the first time in the NEI plasma.
 \end{abstract}

\newpage
\section{Introduction}

X-ray spectra of optically-thin hot plasmas in supernova remnants (SNRs) provide key information on the nucleosynthesis during the stellar  evolution and the supernova (SN) explosion. Since iron (Fe) is the final product of the major nuclear reaction network in a massive star and a SN, it is particularly important element.  So far, Fe abundances of some SNRs have been estimated using the Fe L-shell lines, which are dominant in the low energy band around 1\,keV.  This energy band is complex with other strong emission lines such as oxygen (O) and neon (Ne). Moreover the Fe L-lines consist of many emission lines in this narrow energy band, and hence conventional X-ray detectors such as X-ray CCD cannot resolve these many lines.   
The Fe K-shell lines at 6.4--6.7\,keV are simpler and their emission model is more reliable than those of the L-shell lines. However, due to the limited line fluxes and the sensitivity in the energy band above 6\,keV, the Fe K-shell lines have been reported from only a limited number of X-ray bright SNRs.

Core-collapse (CC) SNe eject less Fe than those of Type Ia SNe (\citealt{Nomoto1984}; \citealt{Iwamoto1999}, and references therein).  Most of the Fe in the center of the progenitor star is collapsed into a central compact object (neutron star or black hole). Thus observationally,  Fe K-shell lines would become weaker than those in Type Ia. The flux of Fe may come from the combined effect of the elemental abundances and the thermal state of the gas.  Thus the differences between Type Ia and CC SNRs may be related to the initial condition of the circumstellar medium (CSM) and an explosion mechanism.

Fe K-shell lines have been reported from other candidates of CC SNRs (e.g., W49B: \citealt{Ozawa2009}; IC443: \citealt{Yamaguchi2009}; G349.7+0.2: \citealt{Slane2002} and G350.1-0.3: \citealt{Gaensler2008}; 3C397: \citealt{Chen1999}). However, some of them are controversial whether they are CC SNRs or not.  The most reliable criteria of CC SN are the presence of a neutron star (pulsar), a pulsar wind nebula (PWN), and oxygen-rich (O-rich) knots, because O is largely enhanced in the ejecta of a CC SN.  At present, three O-rich remnants, Cas A, Puppis\,A and G292.0$+$1.8, have been reported in our Galaxy (see \citealt{Vink2012}, and references therein).  Cas A is the most luminous, and hence the Fe K-shell line was firstly detected at 6.6\,keV (\citealt{Tsunemi1986}). The spatial distribution is not a simple stratified structure (\citealt{Willingale2002}). Thus the SN  explosion would be highly asymmetric.  
Puppis\,A has also the interesting distribution of the ejecta, in that the ejecta is found only in the east, mostly north-east portion (\citealt{Winkler1985}; \citealt{Hwang2008}; \citealt{Katsuda2010}), while a neutron star is propelled in the opposite direction (e.g., \citealt{Becker2012} and references therein).
Such a recoil between SN ejecta and a neutron star is expected in a recent SN explosion model (\citealt{Scheck2006}), and would be examined by asymmetric Fe distribution.  However, no Fe K-shell line has been observed from Puppis\,A.

For the study of the Fe K-shell line in CC SNRs, we observed G292.0$+$1.8. The discovery of a pulsar and a PWN in G292.0$+$1.8 is further confirmation of a CC SNR (\citealt{Hughes2001}, \citeyear{Hughes2003}; \citealt{Camilo2002}; \citealt{Gaensler2003}). The pulsar is located about 0.9\arcmin southeast from the geometrical center of the SNR.  The distance of G292.0$+$1.8 is estimated to be  6\,kpc  (\citealt{Gaensler2003}),  and  the  age  is likely 2990$\pm$60\,years (\citealt{Winkler2009}). 
 
The morphology of G292.0$+$1.8 consists of many small knots and the central belt-like filaments running from the east to the west (\citealt{Park2002}). The central filaments have a normal solar-type composition, suggesting that these are the shocked CSM.  \citet{Lee2010} reproduced the intensity profile of the outer CSM region by a slow wind from a red supergiant (RSG) star with the total mass of the wind of 15--40\,{\it{M}}$_\odot$. The implied progenitor mass ($M >$ 20\,{\it{M}}$_\odot$) was in plausible agreement with previous estimates (\citealt{Hughes1994}; \citealt{Gonzalez2003}; \citealt{Park2004}).
The knots have an enhanced metallicity; Si is enhanced in north-northeast, O is enhanced primarily in southeast, Ne is in northwest and southeast, and Mg is in northwest (\citealt{Park2002}). These knots are probably ejecta origin. The asymmetric distribution of the ejecta elements is interpreted to be non-uniform thermodynamic conditions of the X-ray-emitting ejecta (\citealt{Park2007}).

In spite of these extensive studies, no significant Fe K-shell line has been detected.  \citet{Park2004} proposed that the ejecta are strongly stratified by composition and the reverse shock has not propagated to the Fe rich-zone yet.  However, the X-ray spectra reported so far are faint in the hard band (except for that of PWN), and hence observed lines have been limited up to sulfur (S) K-shell line.  

In this paper, we report the Suzaku discovery of an Fe K-shell line in a high temperature plasma ($kT_e$ = 2--3\,keV) extending to $E$ = 10\,keV.  K-shell lines of argon (Ar) and calcium (Ca) are also reported.  Based on the wide band spectral analysis, we discuss the nature of G292.0$+$1.8. 
We adopt the solar abundances of \citet{Anders1989}. Unless otherwise specified, all errors represent 1$\sigma$ confidence levels.

\section{Observation and Data Reduction}
The Suzaku satellite (\citealt{Mitsuda2007}) observed G292.0$+$1.8 on 2011 July 22-23 (ObsID: 506062010, PI: K. Koyama)
with the X-ray Imaging Spectrometer (XIS, \citealt{Koyama2007}).
The XIS consists of four X-ray CCD cameras placed on the focal plane of  the X-Ray Telescope (XRT).  All four XRTs are co-aligned to image the same region of the sky. The field of view (FOV) of the XIS is \timeform{18'}$\times$\timeform{18'}. Details of Suzaku, the XIS and the XRT are given in \citet{Mitsuda2007}, \citet{Koyama2007} and \citet{Serlemitsos2007}, respectively.
Three of the XIS (XIS0, XIS2, and XIS3) have front-illuminated (FI) CCDs, sensitive in the 0.4--14\,keV energy band, and the other (XIS1) has a back-illuminated (BI) CCD, with high sensitivity down to 0.2\,keV.
XIS2 has been out of function from 2006 November 9 and a small fraction of the XIS0 area has not been available  from 2009 June 23, both due to the damage by micro-meteorites.  

Data reduction and analysis were performed by the HEAsoft version 6.9. The XIS data were processed with the Suzaku pipe-line software version 2.7.
We combined the 3$\times$3 and 5$\times$5 event files. The response functions were generated by using the CALDB 2012-10-09. After removing hot and flickering pixels, we compiled the data using the ASCA-grade 0, 2, 3, 4, and 6 data.
We excluded the data obtained at the South Atlantic Anomaly, during the earth occultation, at the elevation angle from the earth rim below 5$^{\circ}$ (night earth) and 20$^{\circ}$ (day earth). The exposure time after these screenings was 44\,ks.
The spectral resolution has been degraded due to the radiation of cosmic particles 5 years after the launch, and restored  by the spaced-row charge injection (SCI) technique; the charge traps are filled by the artificially injected electrons through CCD readouts. Details of the SCI technique are given in \citet{Nakajima2008} and \citet{Uchiyama2009}.

\section{Analysis and Result}

\subsection{Combined Analysis of SNR and PWN}

\begin{figure*}
   \begin{center}
         \FigureFile(160mm,100mm){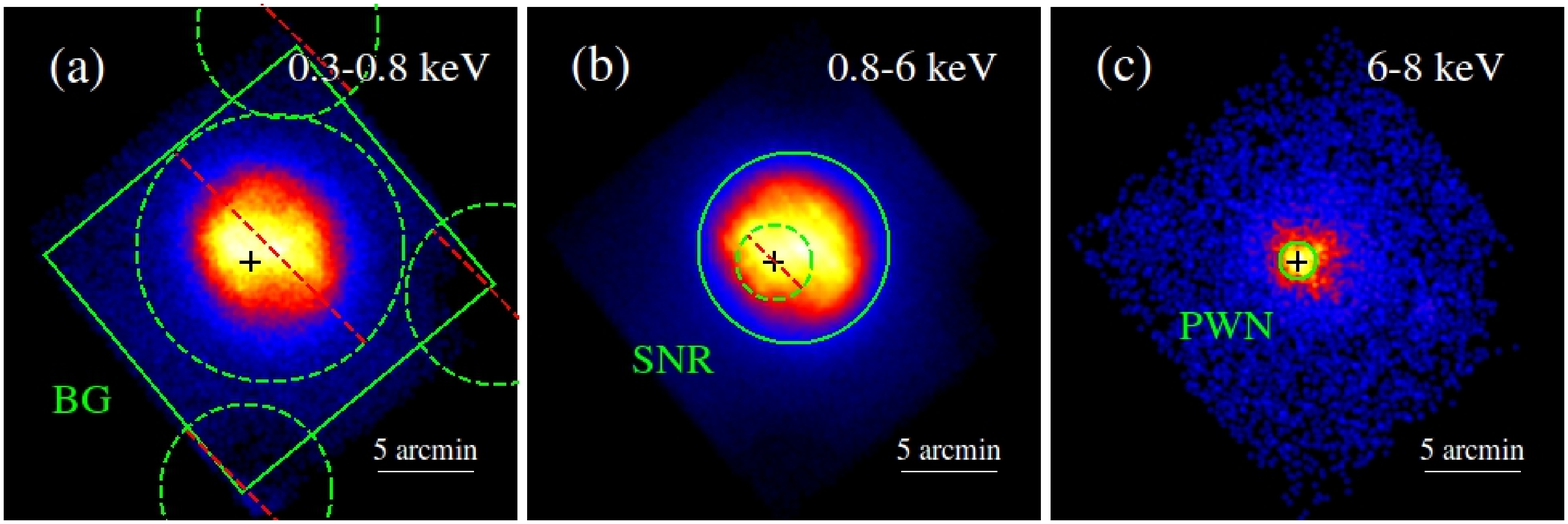}
   \end{center}
   \caption{XIS images of G292.0$+$1.8 in the energy bands of (a) 0.3--0.8 keV, (b) 0.8--6 keV and (c) 6--8 keV, respectively. The geometric center of the PWN (pulsar) is shown with the cross marks. In each image, the spectral extraction regions are shown by the green solid and dashed lines.}\label{xis_image}
\end{figure*}

Figure \ref{xis_image} (a)-(c) shows X-ray images in the 0.3--0.8, 0.8--6 and 6--8\,keV energy bands. In the high energy band above 6\,keV (figure \ref{xis_image} (c)), we see a compact X-ray source at ($\alpha , \delta$) = (\timeform{11h24m39s}, \timeform{-59D16'20''}).  This source corresponds to the pulsar/PWN (\citealt{Hughes2001}, \citeyear{Hughes2003}; \citealt{Camilo2002}; \citealt{Gaensler2003}).
We make two source spectra, one is from the solid circle (radius of \timeform{1'}) as shown figure \ref{xis_image} (c) (here PWN region). The other is from the solid circle (radius of \timeform{5'}) excluding the dashed circle  (radius of \timeform{2'}) as shown in figure \ref{xis_image} (b) (here SNR region).  In figure \ref{xis_image} (a), we show the  background (BG) region: the whole FOV of the XIS (solid square), excluding  the dashed circle (radius of \timeform{7'}). This larger radius than that of the SNR region is employed to avoid the contamination from the SNR. We also exclude the region of calibration sources, which are shown by the dashed circles in the XIS corner. 
For all the regions of PWN, SNR and BG, we separately make the non X-ray background (NXB) spectra using {\it{xisnxbgen}} in the HEAsoft package (\citealt{Tawa2008}).

\begin{figure*}
   \begin{minipage}{0.5\hsize}
   \begin{center}
      \FigureFile(80mm,50mm){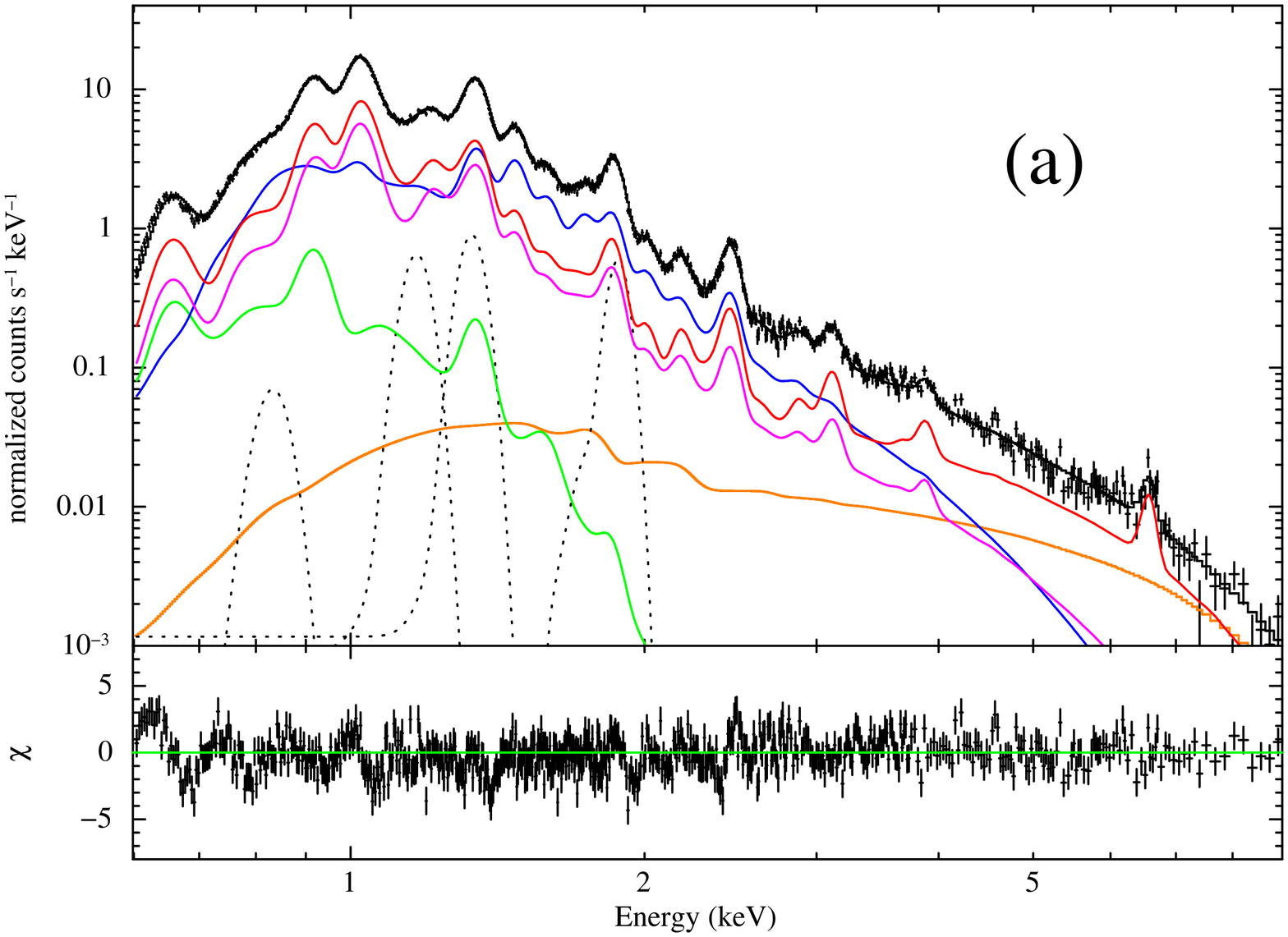}
   \end{center}
   \end{minipage}
   \begin{minipage}{0.5\hsize}
   \begin{center}
      \FigureFile(80mm,50mm){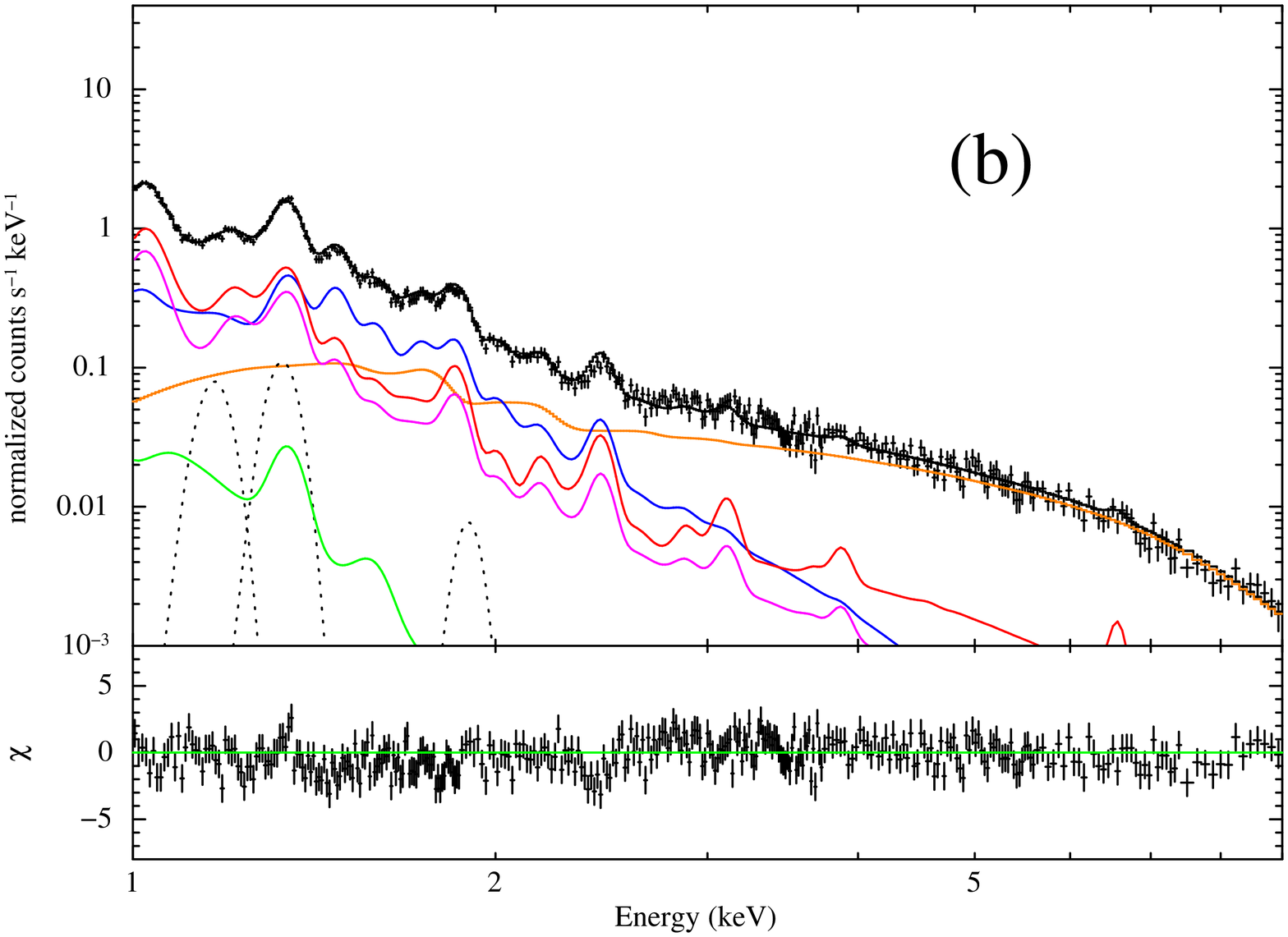}
   \end{center}
   \end{minipage}
      \begin{minipage}{0.5\hsize}
   \begin{center}
      \FigureFile(80mm,50mm){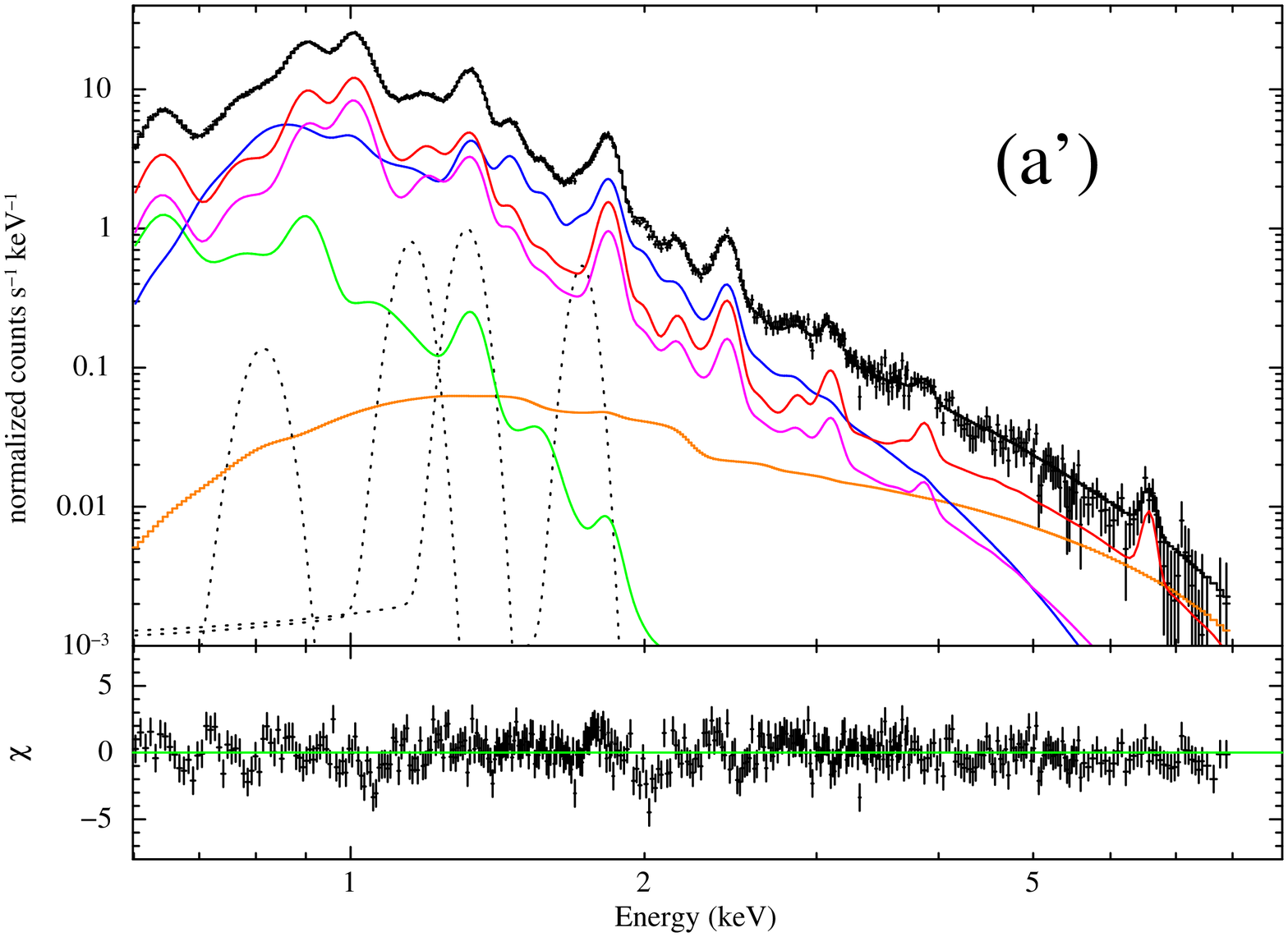}
   \end{center}
   \end{minipage}
   \begin{minipage}{0.5\hsize}
   \begin{center}
      \FigureFile(80mm,50mm){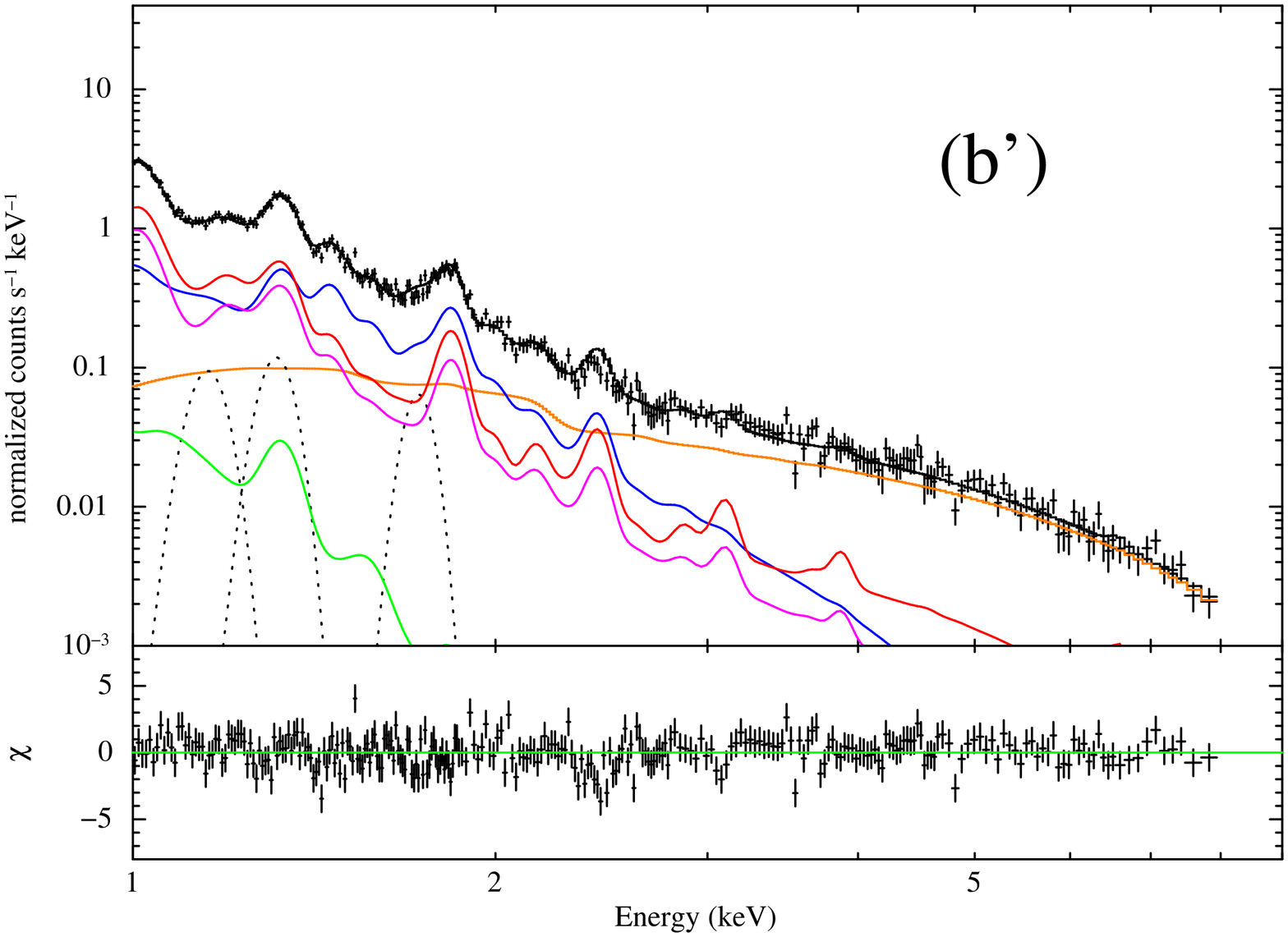}
   \end{center}
   \end{minipage}
   \caption{Spectra of the SNR region and the PWN region.  (a) and (a') are the SNR region spectra for FI and BI, respectively. (b) and (b') are the same panels but for the PWN region. The red and magenta lines are the best-fit 2-VPSHOCK, while the blue and green lines are the best-fit 2-APEC. The orange line and the dotted lines are the power-law and the Gaussian lines. The residuals are shown in the lower panel. }
   \label{SNR-PWN_spec}
\end{figure*}

We make an X-ray background (XB) spectrum from the BG region by subtracting the relevant NXB. The spectra from the SNR and PWN regions are also made by subtraction of relevant NXBs for these regions.  From these spectra, we subtract the XB spectrum  assuming the uniform distribution within the FOV of the XIS after the correction of the vignetting effect. The resulting spectra of the SNR and the PWN regions in figure \ref{SNR-PWN_spec} show many emission lines. From the center energies of these lines, we identify them to be O Ly$\alpha$, Ne He$\alpha$, Ne Ly$\alpha$, Mg He$\alpha$, Mg Ly$\alpha$, Si He$\alpha$ and S He$\alpha$. Thus the spectra should be composed of an optically thin hot plasma with the temperature $kT_e \sim$\,1\,keV (\citealt{Gonzalez2003}; \citealt{Park2004}; \citealt{Lee2010}).  In the spectrum of the SNR region, we find line-like features at 3.1, 3.9 and 6.6 keV, which are likely K-shell transition lines of Ar, Ca and Fe. The Fe line at 6.6 keV is particularly clear.  Therefore, in addition to the 1\,keV plasma (low-$kT_e$ plasma), a higher temperature plasma (high-$kT_e$ plasma) to emit K-shell lines of Ar, Ca and Fe  should be prevailing in the SNR region. We call these two plasmas the SNR components.
The X-rays of the SNR-components  distribute not only in the SNR region but also in the PWN region. On the other hand, according to the observation with Chandra, the PWN is compact of \timeform{30''}--\timeform{45''} (\citealt{Gonzalez2003}), and the spectrum is fitted with a power-law model of $\Gamma\sim$ 1.7 (\citealt{Hughes2001}), indicating its non-thermal nature. We call this power-law emission the PWN-component. The spectrum extracted from the PWN region contains both the SNR-components and the PWN-component. The spectrum from the SNR region, on the other hand, is contaminated by X-rays of the PWN component due to the large point spread function (beam size) of the Suzaku XRT. We hence simultaneously fit the spectra in the SNR and the PWN regions with the combined model, SNR plus PWN-components. The ancillary response files (ARFs) employed in the fit are generated with {\it{xissimarfgen}} (\citealt{Ishisaki2007}). The ARF for the PWN-component  is generated from the Chandra image in the 4--7\,keV band, while that for the SNR-component is made using the thermal emission of the Chandra image (0.6--2.0\,keV), where the emissions of the PWN-component are excluded. 
The energy ranges of the PWN and the SNR regions are 1--9\,keV and 0.6--9\,keV, respectively.  The former energy band is selected  because the contamination of the SNR-component in the PWN region becomes large below 1\,keV. Considering the background level, we also ignore the energy band upper than 9\,keV for the FI, while 8\,keV for the BI.

\citet{Park2004} and \citet{Gonzalez2003} reported that the spectra are significantly different from position to position. Therefore, the integrated spectrum from the entire SNR cannot be described by any single component model fit.
We thus search for a many-components model, starting from one-component model then adding another component one by one, monitoring how much $\chi^2$ is reduced. We use VPSHOCK (\citealt{Borkowski2001}) to represent multi-$n_e t$ non equilibrium ionization (NEI) plasma, where $n_e$ and $t$ are the plasma density and elapsed time after the shock heating. In order to fine-tune the calibration errors, between XIS0, 1 and 3, the gains and normalizations are set to be independent parameters for each XIS.

A 1-VPSHOCK model fails with extreme large $\chi^2 / $d.o.f. of 15006/2071 = 7.25. A 2-VPSHOCK model is largely improved the fitting with $\chi^2 / $d.o.f. of 5400/2060 = 2.62, but still unacceptable. We thus add the third VPSHOCK component (VPSHOCK 1, 2 and 3), then $\chi^2 / $d.o.f. is improved to 5076/2049 = 2.48.  Although the decrease of $\chi^2 / $d.o.f. is only $\delta$ = 0.14, the decrease of $\chi^2$ is 324, which is statistically highly significant. In fact, we check the significance using an F-test tool in the Xspec package, then this process is significant with better than 0.01\% level.  Although errors are large, the best-fit abundances in VPSHOCK 1 and 2 are the same with each other. We hence link the abundances in these two VPSHOCK components. Also abundances of Ni and Ca are linked to Fe and Ar, respectively. Since $n_e t$ of VPSHOCK 3 is 10$^{12-13}$ cm$^{-3}$s$^{-1}$, we replace this model by an APEC model (collisional ionization equilibrium plasma model; CIE). The $\chi^2 / $d.o.f. of this fit is 5184/2062 = 2.51, leaving large residuals in the low energy band. We thus added another APEC component linking the abundances to the APEC component in the 3-component model.
This another APEC improve the $\chi^2 / $d.o.f. to 4838/2060 = 2.35, the F-test statistical significant is even better than the previous process.
Though this $\chi^2 / $d.o.f is still large from a statistical point of view, its value would be due to non-negligible systematic errors. In fact, we find line-like residuals at about 0.82, 1.2, 1.3 and 1.8\,keV for both FI and BI, and 1.7\,keV for BI. The 1.7 and 1.8\,keV line structures are due to the well-known problem of the response function near the neutral Si K-edge energy at 1.84\,keV (\citealt{Yamaguchi2009}). The other line structures would be due to the incompleteness of the VPSHOCK model code. The lines at 0.82, 1.2 and 1.3\,keV correspond to Fe-L complex (\citealt{Uchida2013}; \citealt{Nakashima2013}).  We thus added extra 5 Gaussians to compensate these line-like residuals. The normalization factors of these Gaussians are linked between FI and BI, but that of the 1.7 and 1.8\,keV lines are treated as an independent parameters between FI and BI (\citealt{Suchy2011}). The calibration errors of the contamination  on the optical blocking filter has some problems in the low energy band  (http://www.astro.isas.jaxa.jp/suzaku/doc/suzaku\_td/). For a fine-tuning of the cross errors between the FI and BI CCDs, we allow the $N_{\rm H}$ value in the BI CCD to be independent from the FI CCDs. 
The results of the combined fits by this model are shown in figure \ref{SNR-PWN_spec}. The best-fit parameters are given in table \ref{SNR-PWN_para}. We finally improve the $\chi^2 / $d.o.f. to 2872/2059 = 1.39. The F-test statistical significant is better than 0.01\%.
Thus, we regard that this model (2-VPSHOCK $+$ 2-APEC $+$ PL $+$ 5-Gaussians) is a reasonable approximation for the SNR and PWN spectra, and apply in the following analysis and discussion.

\subsection{Spatial Analysis of SNR}

\begin{figure}
   \begin{center}
         \FigureFile(53mm,53mm){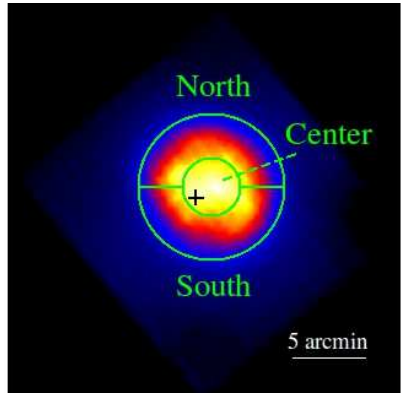}
   \end{center}
   \caption{Region definition for the spatial analysis. Each region is shown in the green solid line. }\label{N-C-S_region}
\end{figure}

\begin{figure*}
   \begin{minipage}{0.5\hsize}
   \begin{center}
      \FigureFile(80mm,50mm){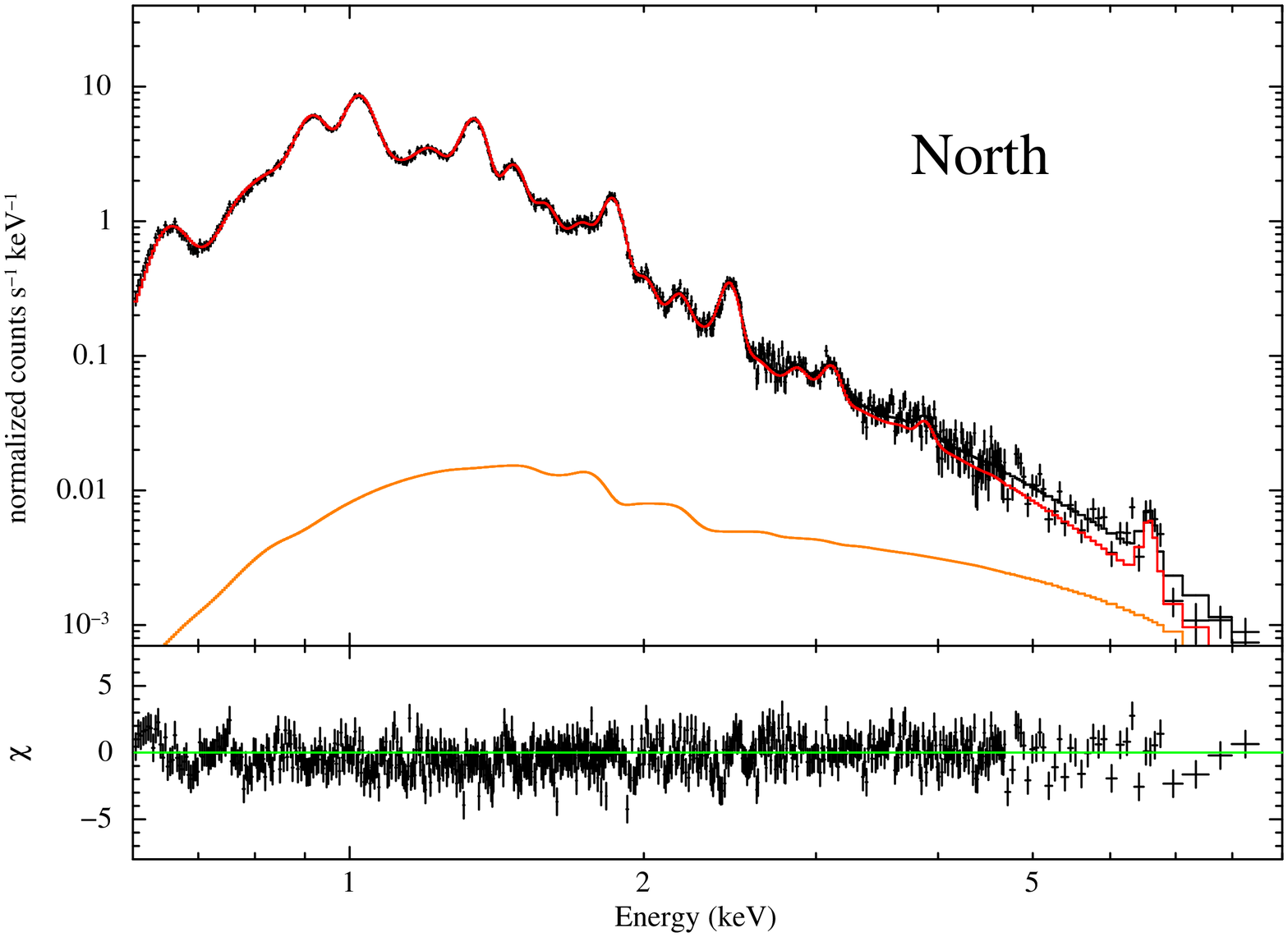}
   \end{center}
   \end{minipage}
   \begin{minipage}{0.5\hsize}
   \begin{center}
      \FigureFile(80mm,50mm){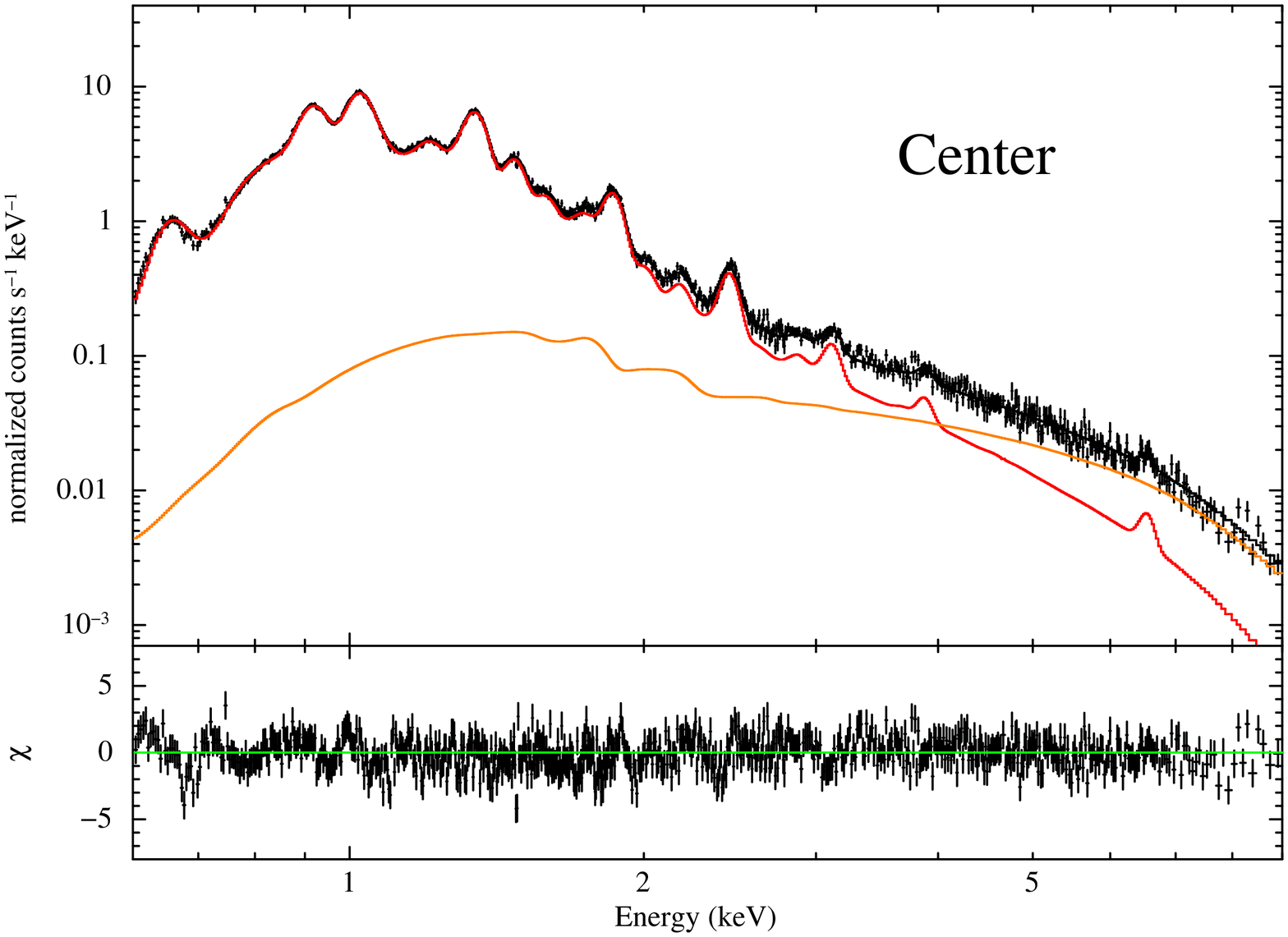}
   \end{center}
   \end{minipage}
   \begin{center}
      \FigureFile(80mm,50mm){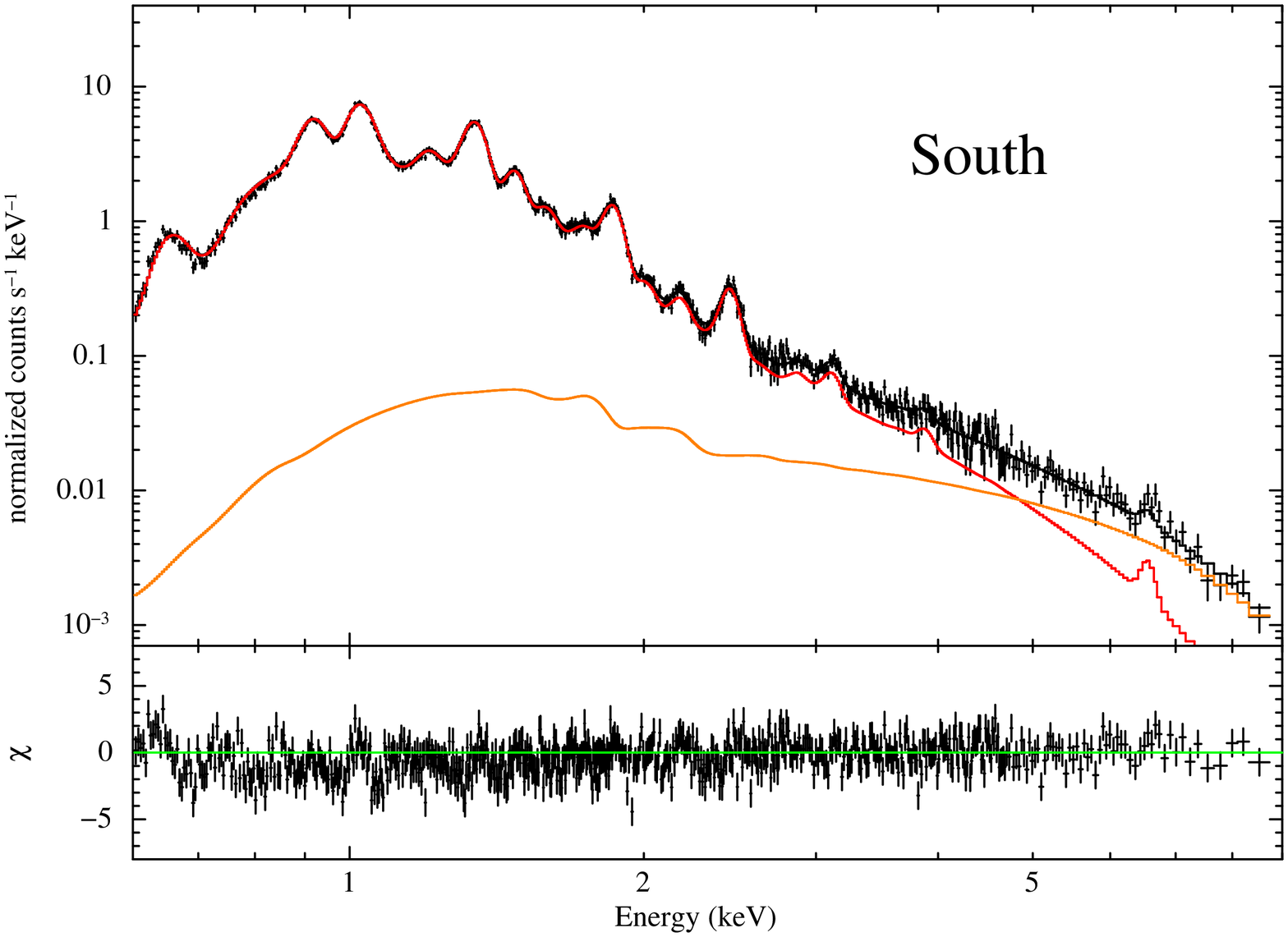}
   \end{center}
   \caption{FI spectra obtained from the regions,  North, Center and South. Each spectrum is fitted with the same model as the combined analysis of SNR and PWN with adding the best-fit PWN spectrum. The red and orange lines represent the SNR and PWN components, respectively.}
   \label{each_region_spec}
\end{figure*}

In order to examine spatial asymmetry of the elements in the SNR, we make spatially-resolved spectra. Since the spatial resolution of Suzaku is limited compared to the size of G292.0$+$1.8 ($\sim$\,\timeform{9'}$\times$\timeform{9'}), we crudely divide the SNR into 3: the center, north and south regions as shown in figure \ref{N-C-S_region} by the solid lines. The spectra are given in figure \ref{each_region_spec}.  We find no significant differences, except a hint of Fe K$\alpha$ flux variations among the 3 regions. For quantitative estimate, we fit with the same model and the method given in subsection 3.1, but $N_{\rm H}$s are fixed to the best-fit values in table \ref{SNR-PWN_para}.  
We obtain nice fit with  $\chi^2 / $d.o.f. of 1283/973 = 1.32, 1391/1055 = 1.32 and 1380/1022 = 1.35, for the north, center and south regions, respectively. 
The best-fit spectra are given in figure \ref{each_region_spec} by the solid histograms.  The best-fit parameters, including abundances, show no spatial-difference within their large errors (typical errors for the abundances are $\sim$50\%).  The only spatial-difference is found in the Fe abundances of the NEI plasma (2-VPSHOCK), which are 1.7$\pm$0.5, 0.75$\pm$0.22 and 1.0$\pm$0.4, for the north, center and south regions, respectively.

The Fe abundance in the NEI plasma may be affected by the flux of the power-law component (PWN component),  because the continuum emission of the ejecta is equal or even less than the power-law emission except the north region (see figure \ref{each_region_spec}).
We therefore re-fit the spatially-resolved spectra, changing the  normalizations of the PWN component by $\pm$5\% of the value in table \ref{SNR-PWN_para},  and fixing the photon index of $\Gamma$ = 1.91. The fit gives no essential changes of the Fe abundances from those of the original value.

\section{Discussion}

In the plasma evolution in SNRs, the X-ray emissions have two different components: the CSM (plus ISM) heated by the blast wave, and the ejecta from the progenitor star heated by the reverse shock.  In the spectral fitting of G292.0$+$1.8, we find two type plasmas in CIE and NEI conditions; the 2-APEC ($kT_e\sim$\,0.2 and 0.7\,keV) and the 2-VPSHOCK ($kT_e\sim$\,1 and 2.5\,keV) plasmas. We call these two type plasmas, the low-$kT_e$ plasma and the high-$kT_e$ plasma, respectively. Since the low-$kT_e$ has nearly solar abundances for all elements and the high-$kT_e$ has super-solar abundances (see table \ref{SNR-PWN_para}), these would be the CSM plus ISM and the ejecta origin of a CC SN, respectively.  
Chandra spectra from many selected regions of bright small spots are described by 1-VPSHOCK model with super-solar abundances (\citealt{Park2004}), while those from the faint outer-most shell are 1 or 2-VPSHOCK model with sub-solar to solar abundances (\citealt{Gonzalez2003}; \citealt{Lee2010}). These results of "no-CIE" plasma are in contrast to the existence of CIE ($\sim$ solar) components in the Suzaku spectra. Since the Chandra results are from selected spots or filaments and those of Suzaku are from the entire SNR region, we suspect that our CIE plasmas are prevailing over the entire SNR except the outer-most shell, while the bright spots are dominated by the VPSHOCK plasma.

We discover Fe K-shell line at 6.6 keV in the eject plasma for the first time. The energy indicates that ionization state of Fe is around B-like.  This medium ionization state is similar to another young CC SNR, Cas A, but is in contrast to nearly Ne-like states in young well known Type Ia SNRs, Tycho, Kepler, and SN\,1006.

\begin{figure}
   \begin{center}
      \FigureFile(80mm,80mm){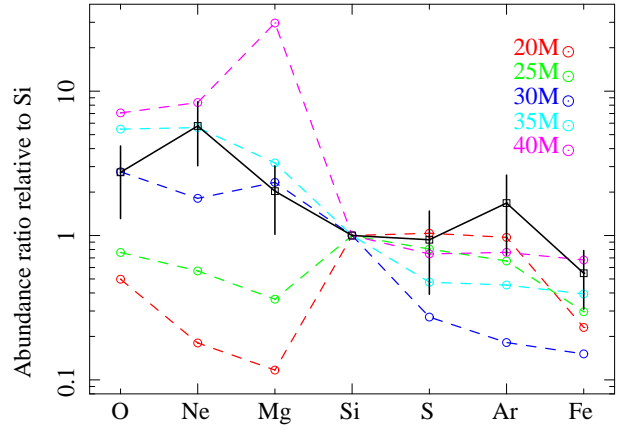}
   \end{center}
   \caption{The solid black line shows abundance ratios of O, Ne, Mg, Si, S, Ar and Fe relative to Si in the high-$kT_e$ plasma (ejecta).
The red, green, blue, light blue, and magenta dashed lines represent core-collapse models with progenitor masses of 20, 25, 30, 35 and 40\,{\it{M}}$_\odot$, respectively (\citealt{Woosley1995}).}\label{ejecta_abundance_pattern}
\end{figure}

Figure \ref{ejecta_abundance_pattern} is the abundance plot of the ejecta for O, Ne, Mg, Si, S, Ar and Fe relative to Si (from table \ref{SNR-PWN_para}) together with  those of the CC SN model in various progenitor masses (\citealt{Woosley1995}). 
We see that the observed abundance pattern is globally in agreement with the model of 30--35\,{\it{M}}$_\odot$.  These mass range of the progenitor star confirm 
the previous report of 30--40\,{\it{M}}$_\odot$ (\citealt{Gonzalez2003}), which was estimated based on the limited information of non-detection of the explosively synthesized heavy elements such as Ar, Ca and  Fe.  
One may argue that CC SN of a massive progenitor 30--35\,{\it{M}}$_\odot$ may leave a black hole rather than a neutron star.  However, other observations show that a neutron star can be still formed from even these massive progenitor stars (\citealt{Muno2006}).

Although \citet{Park2004} and \citet{Gonzalez2003} reported  significant spatial variations in the sub-arcmin scale, we find no significant and systematic variations in the spatial scale over arcmin.  In fact, the best-fit abundances of most of the heavy elements show no variations within their typical errors of 50\%.
Nevertheless, we find marginal evidence of spatial variation of Fe in the ejecta; the north region is enhanced compared to that of the center region.  
Since the position of the neutron star (PWN) is off-set to southeast from the geometrical SNR center (\citealt{Park2007}), it would be conceivable that Fe from the core region would be ejected  to the opposite northwest direction.
Our observational result of the Fe variation is marginal to support this off-set effect due to large errors. To establish this kick-off scenario, we need higher quality observations.

The best-fit spectral parameters of the PWN, the photon index and unabsorbed flux (4--8\,keV) are 1.91$\pm$0.03 and (3.80$\pm$0.18)$\times$10$^{-4}$ photons s$^{-1}$ cm$^{-2}$, respectively. The photon index is steeper than that of the pulsar (1.6--1.7, \citealt{Hughes2001}, \citeyear{Hughes2003}). Probably the index increases as the distance from the central pulsar increases (e.g. \citealt{Slane2000}). 
The PWN flux is 52\% of the total flux (4--8\,keV) from the whole SNR ((7.30$\pm$0.06)$\times$10$^{-4}$ photons s$^{-1}$ cm$^{-2}$). 
This ratio is slightly smaller than 66\%, determined with the high spatial resolution observation of Chandra (Hughes et al. 2001). This difference, however,  would be within uncertainty range due possibly to the NXB and CXB subtraction\footnote{The 4-8 keV band fluxes of the BG region are about 8\% and 70\%  of the whole SNR regions for the data of Suzaku and Chandra, respectively}, and/or other systematic cross errors including different data reduction processes between Suzaku and Chandra. 
Thus our simultaneous fitting analysis properly estimates the flux and spectra of both the SNR and the PWN, although the spatial resolution of Suzaku is limited to completely separate the emissions from these two sources.

\section{Summary}

\noindent We have analyzed Suzaku/XIS data obtained from G292.0$+$1.8. The results are summarized as follows:

\begin{enumerate}
\item  We confirm that the thermal X-ray emission from G292.0$+$1.8 consists of two type plasmas in CIE and NEI conditions.

\item The NEI plasma includes K-shell line from B-like Fe, with super solar abundances for O, Ne, Mg, Si, S, Ar, and Fe.  Therefore this plasma is likely the ejecta origin of the CC SNR.

\item  Using the abundance pattern of the ejecta, we confirm the progenitor mass to be 30--35\,{\it{M}}$_{\odot}$.

\item  The CIE plasma has nearly solar abundances for all the relevant elements, and hence is likely the CSM and ISM origin. 

\end{enumerate}

\section*{Acknowledgments}

We thank all members of the Suzaku operation and calibration teams.
This work is supported by Japan Society for the Promotion of Science (JSPS) KAKENHI Grant Number 24540229 (Katsuji Koyama), 23000004 (Hiroshi Tsunemi), 23340071 (Kiyoshi Hayashida), 24684010 (Hiroshi Nakajima), 12J01194 (Hiroaki Takahashi), 12J01190 (Shutaro Ueda), 24740167 (Koji Mori), 25800119 (Satoru Katsuda), 11J00535 (Hiroyuki Uchida). S.K. is also supported by the Special Postdoctoral Researchers Program in RIKEN.

\newpage
%\bibliographystyle{pasj}
%\bibliography{reference}

\newpage

\begin{table*}[h]
\begin{center}
\caption{Best-fit parameters for the combined analysis of SNR and PWN.}\label{SNR-PWN_para}
   \begin{tabular}{@{\quad}l@{\qquad\qquad}l@{\qquad\qquad}c@{\quad}}
   \hline
   Component & Parameter & Value \\ \hline
   Absorption & $N_{\rm H}$ (10$^{21}$ cm$^{-2}$) & 4.44$\pm$0.19 (FI)\\
                 &         & 4.11$\pm$0.19 (BI)\\
   Power-Law & photon index & 1.91$\pm$0.03 \\
                 & Absorbed flux\footnotemark[$\dagger$] & 3.71$\pm$0.17 \\
                 & Unabsorbed flux\footnotemark[$\dagger$] & 3.80$\pm$0.18 \\
   APEC 1 & $kT_e$ (keV) & 0.17$\pm$0.04\\
                     & O & 0.58$\pm$0.30\\
                     & Ne & 0.74$\pm$0.56\\
                     & Mg & 1.69$\pm$0.27\\
                     & Si & 0.80$\pm$0.16\\
                     & S & 0.83$\pm$0.50\\
                     & Ar (=Ca) & $<$ 1.21\\
                     & Fe (=Ni) & 0.36$\pm$0.09\\
                     & VEM (10$^{11}$ cm$^{-5}$)\footnotemark[$\ddagger$] & 230$\pm$160\\
   APEC 2 & $kT_e$ (keV) & 0.72$\pm$0.01\\
                     & VEM (10$^{11}$ cm$^{-5}$)\footnotemark[$\ddagger$] & 105$\pm$15\\
   VPSHOCK 1 & $kT_e$ (keV) & 1.07$\pm$0.19\\
                     & O & 8.5$\pm$3.5\\
                     & Ne & 17.8$\pm$6.1\\
                     & Mg & 6.3$\pm$2.4\\
                     & Si & 3.1$\pm$1.0\\
                     & S & 2.9$\pm$1.4\\
                     & Ar (=Ca) & 5.2$\pm$2.4\\
                     & Fe (=Ni) & 1.7$\pm$0.5\\
                     & $n_e t$ (10$^{11}$ cm$^{-3}$ s) & 3.0$\pm$2.6\\
                     & VEM (10$^{11}$ cm$^{-5}$)\footnotemark[$\ddagger$] & 6.3$\pm$2.3\\
   VPSHOCK 2 & $kT_e$ (keV) & 2.67$\pm$0.41\\
                     & $n_e t$ (10$^{11}$ cm$^{-3}$ s) & 0.86$\pm$0.19\\
                     & VEM (10$^{11}$ cm$^{-5}$)\footnotemark[$\ddagger$] & 5.2$\pm$2.0\\ \hline
   $\chi^2 / $d.o.f. &  & 1.39 (2872/2059)\\ \hline
   \multicolumn{3}{@{}l@{}}{\hbox to 0pt{\parbox{100mm}{\footnotesize
      \vspace{3mm}
      \textbf{Notes.} Abundances are in units of solar.
      \par\noindent
      \footnotemark[$\dagger$] Flux (10$^{-4}$ photons s$^{-1}$ cm$^{-2}$) in the 4--8 keV band.
      \par\noindent
      \footnotemark[$\ddagger$] Volume emission measure, $\int n_e n_{\rm H} dV / (4 \pi D^2)$, where {\it{V}} and {\it{D}} are the emitting volume (cm$^3$) and the distance to the source (cm), respectively.
    }\hss}}
  \end{tabular}
\end{center}
\end{table*}

\end{document}